\documentstyle[12pt,aasms4]{article}

\lefthead{Neugebauer \& Matthews}
\righthead{10~$\mu$m Quasar Variability}

\begin{document}

\title{Variability of Quasars at 10~$\mu$m}

\author{G. Neugebauer \and K. Matthews}
\affil{Palomar Observatory, California Institute of 
Technology, 320-47,
Pasadena, CA 91125\\
Electronic mail: gxn@caltech.edu, kym@caltech.edu}

\begin{abstract}
Twenty five low redshift quasars have been monitored for several
decades at five near- and mid-infrared wavelengths to detect rapid
variations which would indicate that a nonthermal component was present
in the ``10~$\mu$m bump". Such variability has apparently been detected
in several radio loud quasars and in the radio quiet quasar
PG1535+547.  In addition, the structure function of PG1226+023 shows
that an apparently periodic component is present in its near-infrared
emission.
\end{abstract}
 
\keywords{quasars, infrared, variability}
\newpage

\section {Introduction}

The presence of a local maximum in the continuum spectral energy
distribution of quasars at rest wavelengths around 10~$\mu$m has been
well established (see, e.g., Sanders et al. 1989; Elvis et al.~1994).  
The maximum is
ubiquitous in both radio loud and radio quiet quasars, and the two
types of quasars cannot be distinguished from each other on the basis
of their infrared properties.  The mechanism responsible for the
emission in the ``bump" has  been the subject of debate (see e.g.,
Bregman 1994; Ulrich, Maraschi \& Urry 1997) although the ultimate
source of the luminosity is generally assumed to be the release of
gravitational energy by matter  in an accretion disk falling into a
black hole. Both nonthermal emission from the nucleus and thermal
emission from heated dust is assumed to contribute to the infrared
emission in most quasars, but the infrared emission in radio loud
quasars has generally been taken to be dominated by nonthermal emission
(see e.g., Robson et al. 1993; Bloom et al. 1994 and references
therein). On the other hand, the evidence is strong that thermal
emission from heated dust dominates the emission of radio quiet
quasars  (see the discussion and references in e.g., Sanders et al.;
Bregman).  In particular, the spectral energy distribution around
1~$\mu$m exhibits an  almost universal minimum which is most naturally
explained  by the sublimation of dust grains at temperatures of about
2,000~K.  There is also a drop in the continuum at submillimeter
wavelengths which again is  naturally explained by a drop in the
opacity of dust grains  with increasing wavelength.

As pointed out by Sanders et al. (1989), synchrotron models for the
emission from quasars require brightness temperatures exceeding
m$_e$c$^2$/k$\approx$10$^{10}$~K or about  $10^8$ times larger than the
brightness temperatures which characterize  thermal emission.
Consequently the emitting region, if the emission at 10~$\mu$m is
thermal, will be $\approx$10,000 times larger than that of a nonthermal
source, and therefore the detection of a small source would eventually
be able to settle unambiguously whether nonthermal emission was
present. The predicted angular sizes even from thermally emitting
grains around quasars are, however, typically much smaller than
0.1$''$ and such small sizes cannot be measured directly at the present
time (see  e.g., Sanders et al.).

The extreme size difference implies that a way to discriminate between
the two mechanisms is to look for variability in the mid-infrared
emission. If the emission varies on short time scales, causality argues
that the size of the emitting region is small and thus that the
emission is nonthermal. As discussed below, for a central source with a
typical quasar luminosity, grains which dominate the thermal emission
at the mid-infrared wavelengths are located tens to hundreds of light
years from the nucleus and so significant variability on a time scale
shorter than decades would rule out thermal models.  Note, however,
that a lack of variability is not sufficient  to rule out nonthermal
models.

In this paper we present observations of 25 quasars at 10.6~$\mu$m
intended to distinguish between thermal and nonthermal emission in the
10~$\mu$m bump.  The 10.6~$\mu$m atmospheric window was chosen as the
longest wavelength at which to conveniently make routine measurements
from a ground based site.  Measurements at the mid-infrared wavelengths
such as 10.6~$\mu$m or longer are  necessary since, even if the
radiation is thermal emission from heated dust, the shorter infrared
wavelengths sample hotter grains closer to the nucleus and short term
variations of their thermal re-radiation can exist  as a direct result
of the variable nonthermal emission from the nucleus. The Hubble
constant will be taken to be H$_0$=75~km~s$^{-1}$~Mpc$^{-1}$ unless
otherwise specifically stated.

Observations were also made at wavelengths in the near-infrared
atmospheric windows in order to support the 10.6~$\mu$m measurements.
These observations proved interesting in their own right and thus a
preliminary analysis of these data is included in this paper.

\section {Sample} 

Twenty five quasars from the list of optically selected  quasars of the
Palomar Bright Quasar Survey (Schmidt \& Green 1983) were monitored for
variability in the near-infrared and are tabulated in Table~1.   The
sample was made up of  quasars in the Palomar Bright  Quasar  Survey
which had been established to be bright at 10.6~$\mu$m from
measurements made by the Caltech group (Neugebauer et al. 1987).  The
selection was in  no way systematic or exhaustive; any other bias in
the selection has not been established.

Twenty of the quasars in the sample were detected by IRAS at
60~$\mu$m and most of these quasars were seen in all of the  IRAS
bands (Sanders et al. 1989). Exceptions are PG1535+547 which was
detected  by IRAS only at 12 and 25~$\mu$m, not at 60 and
100~$\mu$m, and the quasars PG0026+129, PG1617+175 and PG2302+029
which were not detected in any of the IRAS bands. PG2209+184 is
in a location which was not scanned by the IRAS satellite.

The spectral energy distributions from 0.3 nm to 6 cm are shown
in Sanders et al.~(1989)  while the spectral energy distributions
of thirteen quasars of the sample from 10 keV x-rays to radio
wavelengths are given by Elvis et al.~(1994). Wilkes et
al.~(1998) include the quasar PG1351+640 in their study of ISO data of 
the far-infrared continuum of quasars.

Bolometric luminosities,  which represent the estimated luminosity of
the quasars from $10^{12}$ to $10^{17}$ Hz, are taken from Sanders et
al. (1989).  They  are included in Table~1 and their distribution is
shown in Figure~1; the median bolometric luminosity is
$L_{bol}$=1.6$\times 10^{12}$$ L_{\sun}$.  Schmidt \& Green (1983) have
designated five of the objects in the sample (PG0007+106, PG1119+120,
PG1501+106, PG1535+547 and PG2209+184) as ``Seyfert I nuclei or
low-luminosity quasars" based on their having absolute B magnitudes
M$_B > $~-23.0~mag with H$_0$=50~km~s$^{-1}$~Mpc$^{-1}$.  The redshifts
of quasars in the sample are also taken from Sanders et al. and given
in Table~1. The distribution of redshifts is included in Figure~1; the
median redshift is z=0.089.  The mean 10.6~$\mu$m magnitudes obtained
in this study  range from 8$>$[10.6]$_{avg}>$5~mag.

Twenty of the quasars are radio quiet  while five are radio loud; see
Table~1.  The classification as either radio loud or radio quiet is
that given by Sanders et al. (1989).  The separation between the two
types has been made at log[$\nu f_\nu$(6 cm)/$\nu f_\nu$(1~$\mu$m)] $=$
-4 using  the radio data of Kellermann et al. (1994).  The distinction
between radio loud and radio quiet is open to discussion.  For example,
Kukula et al. (1998) have called PG0007+106, which Sanders et al.
classify as radio loud, a ``radio-intermediate quasar". Becker (1998,
private communication) has argued that modern complete surveys at radio
frequencies show there is a continuun between radio loud and radio
quiet quasars.

\section {Observations}

Caltech measurements of selected quasars at 10.6~$\mu$m  were obtained
from 1974 to the 1998. Systematic monitoring observations specifically
to measure variability at 10.6~$\mu$m were started  in
1988.  The observations after that time were made at  the Cassegrain
f/70 focus of the 200-inch Telescope at Palomar Mountain using beams
with 5$''$ diameters.  
Measurements were made in the $J,  H, K, L'$, and $N$ photometric
bands; the central wavelengths and  half power
wavelength widths of the filters used are given in Table 2.  
A single element Ge:Ga bolometer at pumped liquid helium
temperature was used to measure the flux at 10.6~$\mu$m and a single
element InSb photovoltaic detector at solid nitrogen temperature was
used  to observe wavelengths between 1.27 and 3.7~$\mu$m. The  effects
of the sky emission were suppressed by chopping the beam in a
north-south direction with the f/70 secondary mirror at either 5 or 50
Hz and nodding the  telescope, also in a north-south direction, by
either 15$''$ (1.27-3.7~$\mu$m) or 6$''$  (10.6~$\mu$m).  The telescope
tracking was maintained by using an offset guider with a visual guide
star.

For completeness, compatible  observations made before 1988 are
incorporated into the data-base of this study. A few  of these were
made at the 100-inch Hooker Telescope at  Mt. Wilson. During the total
time span, the instrumentation  evolved gradually. The bolometer system
for the $10.6~\mu$m observations has, however, remained intact since
1973 and the same filters have been used in various near-infrared InSb
systems which were introduced in 1972, 1975, 1981 and 1982. Care was
taken to ensure that the photometric results were consistent during all
the changes.

In the decade following 1988, the majority of the 25 quasars in
the sample were sampled once per year, but for a few objects the
time interval between observations sometimes exceeded three
years; all 25 quasars have measurements made at all five
wavelengths on at least seven nights. Observations at the four
shorter wavelengths typically took a total of 20 minutes and those at
10.6~$\mu$m  typically took 40 minutes. Observations were
generally made when the quasars were at $\lesssim$1.7 air masses
and standard air mass corrections were applied. Only observations
made on photometrically clear nights when the full width at half
maximum (FWHM) of the visual image was $\lesssim 2''$ were
included in the final data set. In total there were some 1,900
photometric measurements  of quasars included.

The quasar observations were accompanied by a comparable  number
of measurements of ``standard" stars to maintain  photometric
accuracy. In the four shorter wavelength bands, the stars
identified by Elias et al. (1982) were used. At 10.6~$\mu$m,
stars previously measured by the Caltech  group and tabulated in
Table~3 were utilized. The integrity of the  photometric
standards was checked by the internal consistency of the
photometric sensitivity of several measurements made on the same
night and throughout this study. The (population) standard
deviation of the sensitivity measured on any one night for the
three  shortest bands was typically 0.02~mag; that of the
3.7~$\mu$m-band  was 0.04~mag and that of the 10.6~$\mu$m-band
was 0.05~mag. The  standard deviations of the absolute sensitivities 
of the system in the
different  wavelength bands, i.e. the sensitivity which would
have been assumed if no ``standard" stars were measured, were 
 $\sim$0.1~mag including one
period during 1991-1992 when the sensitivity  of the
different wavelength bands differed by as much as $\sim$0.5~mag.
The sense of the dispersion in these sensitivities was completely
consistent with effects of the mirror coating and from the
eruption of Mt.  Pinatubo.

The uncertainties in the magnitudes were calculated as the square root
of the quadratic sum of the statistical uncertainty in the object plus
the dispersion in the ``standards" plus an uncertainty which is
representative of the systematic uncertainty in making the
observation.  The later was somewhat arbitrarily assigned the value
0.02~mag, based on experience with similar measurements of objects with
large signal to noise ratios.  Generally the statistical uncertainties
dominated the 10.6~$\mu$m observations; systematic uncertainties often
dominated the 1.27 to 2.23~$\mu$m observations.

\section {Results}

Four representative data sets are shown in Figures~2--5; they will be 
discussed in detail below. The data of
the remaining 21 quasars in the sample are given in Figures~13--33 in 
the Appendix.
All these figures are similar in that they include observations  in the
five wavelength bands presented in the same order from bottom to top.
Unless specifically indicated, the  ordinates of each figure are scaled
to be the same for all wavelength bands for that object; the scales
were set to include the largest excursions measured in any band for
that object. Some observations made before 1989 have previously been
published by Neugebauer et al. (1989).

A measure of  the variability relative to the uncertainties
in the measurements for each wavelength band is provided by the
reduced chi-square:
\begin{center}
 
$\chi^2 \equiv \frac{1}{N}\times \sum _{i=1}^{N} \frac{(m_i - <m>)^2
}{\delta m_i ^2}$

\end{center}
   
\noindent 
where  $m_i$ is the magnitude observed on a specific date and
wavelength, $<m>$ is the magnitude of the observations at that
wavelength as determined by  a linear fit to all the measurements at 
that wavelength throughout the study, N is the number of observations and
$\delta m_i$ is the uncertainty in the measurement  of $m_i$. As
discussed below, a long term drift in a nonthermal component can be
transferred to a thermal component. The reduced $\chi^2$ for each 
of the 25 objects in
the sample for each wavelength is included in Figures~2--5 and 13--33.
A limit on the inherent variability necessary to produce large values
of chi-square at 10.6~$\mu$m was set by the typical uncertainty  of
0.1~mag, or 10\% at that wavelength. This contrasts to the shorter
wavelengths where uncertainties of a few percent were routinely
achieved.  Nonetheless, of four of the five radio loud quasars
observed, all but PG1211+143, and six of the 15 radio quiet quasars had
reduced $\chi^2 >$ 1.5 at 10.6~$\mu$m indicating measurable
variability.  At 2.23~$\mu$m, the flux variations from all the radio
loud quasars and all but two radio quiet quasars exceeded this limit.
In fact, as discussed below, the median variability of the majority of
the sample exceeded 0.1~mag at all wavelengths. Thus, we conclude that
we have measured variability, certainly at the shorter wavelengths, but
also at 10.6~$\mu$m, in some of the quasars of this sample.

It is also possible to calculate reduced $\chi^2$ by converting each
magnitude to a flux density, a linear quantity, finding the average
flux density, and then calculating the reduced $\chi^2$.  Although the
reduced $\chi^2$ found in this manner differs slightly from that found
from the magnitudes because of the transformation of the uncertainties,
the two formulations give qualitatively the same results.  The $\chi^2$
derived from the magnitudes and presented with the figures are
generally smaller than those derived using the flux densities.

In some cases, even where the reduced $\chi^2$ is consistent with a
constant slope, there is apparently often a low frequency variation
present with a period of about a decade which appears as a correlated
variation at more than one wavelength.  A different way of presenting
the results, which emphasizes possible correlations in the magnitudes
at the cost of time knowledge, is to compare the photometric changes of
an object that was measured at the same two wavelengths on two nights.
In the following, all possible comparisons, i.e., differences between
all nights when the same two wavelengths were observed, are included.
Examples which illustrate apparent correlated changes at all
wavelengths are shown in Figure~6 (PG1226+023, radio loud) and Figure~7
(PG1535+547, radio quiet) while Figure~8 gives an example of both
corellated and un-correlated changes (PG1351+640, radio loud). Similar
plots show that the changes at 1.27, 1.65 and 2.23~$\mu$m for all the
quasars but PG1700+518 are apparently correlated. On the other hand, 
two thirds of the sample showed variations at
3.7 and 10.6~$\mu$m which are probably not correlated as
evidenced by a probability of the null hypothesis of no
correlation in excess of 0.10; undoubtedly, this is partially the
result of the larger uncertainties at these wavelengths.

\section {Discussion}
\subsection {Thermal model}

In order to  make a comparison with the observed results, a simple
model of thermal re-radiation from heated grains was constructed. Dust
grains were assumed to exist in concentric shells with a range of radii
around a central luminosity source.  The temperature of the grains   at
each radius was set by radiation balance between the photons absorbed
from the central source and re-emitted by  the grains. Silicate,
graphite and blackbody grains were tried. The aggregate of the grains
was assumed to be optically thin. The maximum grain temperature allowed
before the grains were assumed to sublime was 2,000~K for graphite
grains  and 1,500~K for silicate grains.  The minimum temperature was
arbitrarily assumed to be 100~K; the results were insensitive to these
choices.  Planck averaged emissivities were taken from  Draine \& Lee
(1984) and the grain size distribution  was taken to be that given by
Mathis, Rumpl \& Nordsieck (1977).  The grain number  density with size
was assumed to have a power law dependence on radius;  an inverse
square dependence was most often used.

If the  thermal model applies, if the grains are silicates and if the
central bolometric luminosity is the median bolometric luminosity of
the sample, $L_{bol}$=1.6$\times 10^{12}$$L_{\sun}$, the shell  which
contributes the maximum 10.6~$\mu$m signal has a  diameter of about 55
light years. Half the flux observed in  the 10.6~$\mu$m band would be
emitted by grains located in  shells with radii between 15 and 60
light years.  In contrast, if the bolometric luminosity  of the central
source is only 2.8$\times 10^{11}L_{\sun}$, that of PG1535+547 the
least luminous quasar in the sample, in this simplified model the shell
which contributes the maximum 10.6~$\mu$m signal  has a radius of
only about 20 light years.

In order to characterize the response in thermal emission to variations
in the nonthermal  central source, a step increase in luminosity of the
central source was assumed.  The grains  were assumed to respond
instantly to the absorbed energy upon arrival of the photons from the
central source. Typical responses predicted to be  observed at 2.23, 3.7
and 10.6~$\mu$m if the grains were  silicates are shown in the top of
Figure~9 for a step amplitude in the central source from $1.4\times
10^{12}$ to 1.8$\times 10^{12}L_{\sun}$. As expected, the  observed
increase at 2.23~$\mu$m is predicted to be significantly faster  than
that at 10.6~$\mu$m since the grains responsible for  the short
wavelength radiation are hotter and hence closer to  the central
source. Although Figure~9 (top) illustrates the results of the model
only for silicate grains, the amplitude of the response at 10.6~$\mu$m
to a step change in the central  luminosity is about one-third that at
2.23~$\mu$m over a wide range in luminosities and for graphite as well
as silicate grains.

It is difficult from the presentation of Figure~9 to visualize the time
it takes for the grains in the envelope to respond to a change in the
central luminosity. This time can conveniently be characterized by the
time it takes the observed flux to go from 10 to 90\% of its predicted
final level if the central luminosity source has  a step increase in
luminosity. This 10 -- 90\% rise time is insensitive to the step
amplitude,  but depends strongly on the average luminosity of the
central  source.  Figure~9 (bottom) shows the  10 -- 90\% rise times 
expected in
observations at 2.23, 3.7 and 10.6~$\mu$m if the central luminosity
source increases by 25\%.

\subsection {Specific examples}

Figure~2 shows the most extensive set of observations obtained, those
of  PG1226+023 (3C273), a bright radio loud quasar with a flat radio
spectrum. Variability at 10~$\mu$m accompanying variations  over a wide
range of wavelengths has previously been reported  in this quasar by
Rieke \& Lebofsky (1979), Robson et al. (1983), Cutri et al. (1985),
and Courvoisier et al. (1988).  As discussed in these references, the
nonthermal component is apparently sufficiently strong relative to the
thermal component to be evident at all the infrared wavelengths.
Indeed, the present observations confirm variability in PG1226+023 at all
five of the near-infrared and mid-infrared wavelengths including
10~$\mu$m.  PG1226+023 is one of two quasars included in both this
sample and in the study by Smith et al.  (1993), in the Johnson B band,
of a large sample of radio loud quasars.  This study showed apparently
random vaiations in the visual flux with a peak-to-peak amplitude of
about 0.5~mag.

Figure~3 shows observations of PG2209+184, another  radio loud quasar
with a flat radio spectrum; it is conjectured that the effects of  a
non-varying host galaxy masks the variations at wavelengths from 1.27
to 2.23~$\mu$m.  Host galaxies have been observed associated with low
redshift quasars by a large number of authors; see e.g.  Bahcall et
al.  (1997) and references therein. They have been observed in the
near-infrared by, among others, McLeod \& Rieke (1994a, 1994b); recent
near-infrared HST obsevations of host galaxies are presented by Hines
et al (1999; private communications).  Near-infrared images of 15 of
the quasars in this sample that dilineate the host galaxies have been
obtained in an accompanying observing program.  The median contribution
of the putative host galaxy above that of a point like source between
the seeing disk of $<$1$''$ diameter and an outer diameter of 5$''$
(the beam diameter used in this study) is about 10\% of the 1.65~$\mu$m
flux of the quasar. Of those imaged, PG2209+184 has the largest host
contribution to the observed flux; at 1.65~$\mu$m about one quarter of
the quasar flux in a 5$''$ diameter beam  can be attributed to a host
galaxy.  The hosts are thought to be normal galaxies (see e.g.,
Neugebauer, Matthews \& Armus 1995) so their contribution to the
10.6~$\mu$m, and probably 3.7~$\mu$m, flux is negligible, although the
presence of a host galaxy serves to damp out 2.23~$\mu$m and shorter
wavelength variations.  The detailed nature of the hosts will be
discussed in a future publication, but for the purposes of this paper,
their effect can be ignored at 3.7 and 10.6~$\mu$m.

The observations of PG1501+106, a radio quiet quasar, provide a data
set (Figure~4)  where the 10.6~$\mu$m  variability is markedly less
than, and uncorrelated with, that at the shorter wavelengths.
Presumably this is a case where the 10.6~$\mu$m emission is dominated
by thermal emission from dust grains with subsequently smaller
variations at 10.6~$\mu$m  than at the shorter wavelengths.
Specifically, the smooth rise at 10.6~$\mu$m is consistent with the
model described above if there were a long period increase in the
nonthermal near-infrared source in the decade before these observations
were made.

In contrast to the previous example, PG1535+547,  another radio
quiet quasar, shows apparent variability at all  measured
wavelengths as illustrated in Figure~5. Although the variation at
10.6~$\mu$m has a reduced~$\chi^2$   consistent with a constant
flux (reduced~$\chi^2$~=~1.27), the deviations at all of the
wavelengths are apparently correlated (Figure~7)
and thus convince us that true variations on a time scale as
short as decades is being observed at 10.6~$\mu$m. At first
sight, the data in Figure~5 indicate that, in this quasar at
least, nonthermal processes play a significant part in producing
the $10.6~\mu$m bump. It should be noted that PG1535+547 is the
least luminous of the quasars in the sample. Still, the short
time of the variations would require a special composition of the
grains if the radiation were purely thermal. The amplitude of the
10.6~$\mu$m observations in relation to the variations in the
near-infrared is also inconsistent with the simple model of
thermal emission. Neither a thermal or nonthermal model easily
explains the apparent phase advance at 3.7~$\mu$m.

\subsection {Median Variability}

Netzer et al. (1996) have monitored 44 radio loud quasars at B
and R for a period of six years with a mean sampling of four to 
six times per year.  Their sample overlapped the
present one in only one quasar, PG1704+518 (3C351; Figure 31).
This quasar shows a variation in B of $\sim $0.5 mag which is
reflected in a comparable variation in the near-infrared
wavelengths.  All the quasars in Netzer et al's sample varied
during their six year survey, some by as much as 3 mag (maximum
variation) but half of them by 0.2 to 0.4 mag. Presumably, the
variations found in Netzer et al's observations represent an
indication of the nonthermal fluctuations present in a sample of
nonthermal quasars and as such is interesting to compare with the
present survey.
  
As a robust measure of the variability, Netzer et al. (1996)
tabulated the ''median variability" of their survey. The
''median variability" is defined as the median of the absolute
value of the peak to peak magnitude difference taken over the run
of all measurements of a quasar in a particular wavelength band.
The distributions of the median variability  provide a convenient
tool to compare a presumably nonthermal population with the
infrared sample.

Figure~10 (top) shows the distributions of the median variability of
the radio loud and  radio quiet sub-samples as observed in the present 
survey at 10.6~$\mu$m.  The distributions of
the median variabilities of the shorter wavelength near-infrared
observations agree qualitatively with those shown in Figure~10.  
Although there is a slight tendency  at 10.6~$\mu$m for a
higher median variability among the radio loud quasars than among 
the radio quiet quasars,  there is
essentially no difference between the two groups. A further comparison
is made in Figure~10 (bottom) which shows the median variability of the
present 10.6~$\mu$m measurements as  compared with that of the median
variability obtained by  Netzer et al. (1996) in the $B$-wavelength
band (0.44~$\mu$m). Although there is no significant overlap between 
Netzer et al.'s sample and the
present sample, and this median  does not account for the larger
uncertainties in the 10.6~$\mu$m observations or the difference in
sampling frequency, there is no
strong difference between the two  distributions.  
In particular, the medians have comparable amplitudes, and the
infrared variations do not show the decreased amplitude expected
from thermal emission on the simple model described above. 

The variations observed by Netzer et al. (1996) are almost certainly
nonthermal.  Thus we take the similarities present in Figures~10 as
evidence that a nonthermal component is present in the 10.6~$\mu$m
emission of both radio loud and radio quiet quasars.

\subsection {Structure function}

Structure function analysis is useful in discussing variability of
unevenly sampled data sets. The structure function is discussed in the
context of astronomical variability by e.g., Simonetti, Cordes \&
Heeschen (1985), Hughes, Aller \& Aller (1992), Press, Rybicki \&
Hewitt (1992) and Smith et al. (1993).  If the observations  are made
at time intervals which are short compared to true variations in the
quasar flux,  the structure function for short time lags reflects the
measurement noise in the sample.  The structure function for large lags
characterizes the low frequency variations in the quasar flux.

The following discussion closely follows that of Press et al.  (1992).
At each wavelength, a lag $\tau_{i,j}$ was computed for each pair of
observations:

\begin{center}
 
$\tau_{i,j} = \mid\tau_i - \tau_j\mid $
 
\end{center}
\noindent
A one-point estimate of the structure function, $sf_{i,j}$,  was
calculated for each lag:

\begin{center}
 
$sf_{i,j} \equiv{(m_i - m_j)^2 - \epsilon_i^2 - \epsilon_j^2} $
 
\end{center}

\noindent  
where  $m_i$ and $m_j$ are the magnitudes observed at
a specific wavelength at times $\tau_i$ and $\tau_j$, and
$\epsilon_i$ and $\epsilon_j$ are the quoted standard deviations
of the measurement. The structure function estimates were sorted
by their lags, and then binned so that each bin contained at
least 50 individual estimates and lags. The lags and the one
point estimates in each bin were finally averaged.  These
averages for the 10.6~$\mu$m observations  of all the 25 quasars
in the sample are shown in Figure~11.  The figure indicates a
difference in the behavior of radio loud quasars and radio quiet
quasars although the large amplitude  around two  to three years in the
radio loud quasars is mainly the result of one
quasar (PG2209+184).

Although the structure function analysis does not directly
address the question of the nature of the emission around
10~$\mu$m from quasars, the structure function does give clues as
to the nature of the near-infrared emission from PG1226+023, the
best studied of the quasars. The structure functions for
PG1226+023 at all the five wavelengths are shown in Figure 12.
At the three shorter wavelengths, the structure function clearly 
shows peaks at lag times separated by about a decade.  As
a check that these peaks are not an artifact of the processing, a
similar analysis to that described above was carried out on data
which were artificially scrambled in time, but maintained the
sampling pattern and  the intensity levels of the original
observations.  As a result of these checks, we conclude that
the processing is not responsible for the variations observed and
that the structure functions reflect a real property of the
quasar.

As pointed out by e.g., Smith et al. (1993), a sinusoidal variation in
the fluxes gives a sinusoidal variation of similar period in the
structure function such as seen in Figure~12. Thus, the almost 
periodic appearance of the
structure function indicates a quasi-periodic fluctuation in the near-
infrared emission of the quasar. The maximum near ten years in the lag
indicates that an oscillatory  mechanism with a period near a decade is
present in the nonthermal central engine. A maximum was often seen in
the visible structure functions studied  by Smith et al. at lags
between five and ten years and  a double humped profile with a period
roughly a decade was observed in the quasar 1318+290 leading to the
speculation that the  ``periodic" behavior is a feature of the
nonthermal engine of the radio loud quasars. Smith et al., however, do
not call attention to any striking behavior in 
PG1226+023 and thus the
behavior seen in Figure~12 may be restricted to emission in a rather
small wavelength interval. 

The sinusoidal behaviour appears to be  missing from the
3.7~$\mu$m observations of PG1226+023. Numerical simulations of
the data were made where a sine wave with an amplitude of 0.5~mag and a
period of 11 years was sampled as the real data were, given
random deviations proportional to the actual uncertainties and
assigned the uncertainties of the real data. These simulations were able 
to reproduce the data and structure functions at all the measured
wavelengths, including 3.7 and 10.6~$\mu$m, quite realistically.  Thus we
conclude that the absence of an obvious signature can be a result
of the sampling and uncertainties in the data and that the
observations are consistent with an underlying sinusoidal
variation in this quasar at all the observed wavelengths.

The roughly periodic nature of the structure function  present
for PG1226+023 is not a universal feature in all the quasars or
even all the radio loud quasars. As indicated by 
Figure~11, however, the maxima near a decade seem to be present
more often among the near-infrared observations of the radio loud
than among the radio quiet quasars.   

\section {Conclusions} 
There is no single set of observations which, on the basis of their
temporal variations, can unambiguously demonstrate that nonthermal
emission dominates the 10~$\mu$m peak of any
radio quiet quasar.  The strongest case is that of PG1545+47 which
shows apparently correlated variations on time scales less than
decades with similar amplitudes at all the infrared wavelengths. 
In  this object, at least, the most simplistic model of purely
thermal emission cannot unambiguously explain the observations.
Unfortunately, PG1545+47 is the lowest luminosity quasar of the sample
and the predicted lack of variability in thermal emission from dust
around quasars  depends strongly on the luminosity. It is also always
possible, indeed almost certainly true, that special geometries or
constituents, coupled with the low luminosity, can conspire to allow
thermal emission to vary on time scales less than a decade.  The
similarity of the median variability of the 10.6~$\mu$m observations
and that of the near-infrared observations and the visible observations
of Netzer et al. (1996) (Figure 10) argues that some of the 10.6~$\mu$m
radiation is nonthermal.  Thus some nonthermal emission is apparently
present in some radio quiet quasars, i.e., both thermal radiation from
dust and nonthermal radiation contribute to the ``infrared bump".

\section{Acknowledgements}
A project of several decades duration cannot be contemplated without
the support and help of a large number of people.  We thank Lee Armus,
Eric Becklin, Jay Elias, James Graham, Todd Hunter, Jon Kawamora ,
David Shupe, Tom Soifer and Alycia Weinberger as well as
 the night assistants Rick Burruss, Juan Carrasco, Gene Hancock, Skip
Staples and Gary Tuton and the entire staffs of Palomar and Mt. Wilson
Observatories for help obtaining these data. We acknowledge helpful
discussions with Lee Armus, Peter Barthel, Roger Blandford, David Hogg,
Marcia Neugebauer, Sterl Phinney, Annila Sargent, Tom Soifer and Alycia
Weinberger. Caltech, NSF and NASA have supported us over the years.

\newpage
\appendix
\section{Appexdix}
The observations of the 21 quasars not shown in Figures~2--5 are given
in Figures~13--33 in the format of Figure~2.

\newpage
\figcaption[neugebauer.fig1.ps]{(left) A histogram of the 
bolometric luminosities, as described in the text, of the quasars in 
the sample is shown. (right) A histogram of the redshifts of the 
quasars in the sample is given. The
redshifts and bolometric luminosities are from Sanders et al. (1989).}

\figcaption[neugebauer.fig2.ps]{The observations of the quasar PG1226+023 
at (from the
bottom to the top)  1.27, 1.65, 2.23, 3.7 and 10.6~$\mu$m are given as
magnitude differences from the long term weighted average of the 
magnitudes measured in this study at that wavelengh. The ordinate range, 
which is the same at each 
wavelength, is adjusted to include the largest excursions in any band 
and is labeled in the bottom, middle and top panels. The dashed lines 
indicate a weghted linear fit to all the magnitudes obtained in this 
study. For each band, the 
long term weighted average  magnitude determined in this
study and the reduced $\chi^2 $ are given to the right of the figure.}

\figcaption[neugebauer.fig3.ps]{ The same as Figure 2, but for PG2209+184.}
\figcaption[neugebauer.fig4.ps]{ The same as Figure 2, but for PG1501+106.}
\figcaption[neugebauer.fig5.ps]{ The same as Figure 2, but for PG1535+547.}
\figcaption[neugebauer.fig6.ps]{ The magnitude change at one wavelength 
versus that at another wavelength is shown for 
the observations of PG1226+023. The differences in the observations at 
the two wavelengths given in the upper left of each panel (upper---ordinate, 
lower---abscissa) are compared. Solid dots indicate that the 
time interval between measurements was less than five years, 
while crosses indicate longer intervals. Only those cases where
subsequent observations were made on the same nights at two
wavelengths are presented so all the data taken are not included,
and so these data are  more concentrated to the time since 1988. In
those cases where the changes are apparently correlated, formally
where the significance level at which the null hypothesis of zero
correlation is disproved is less than 5\% (see e.g., Press et al. 1985), a
dashed line indicating  the best fit to the correlation is shown.}
\figcaption[neugebauer.fig7.ps]{ The same as Figure 6, but for PG1535+547.}
\figcaption[neugebauer.fig8.ps]{ The same as Figure 6, but for PG1351+640.}
\figcaption[neugebauer.fig9.ps]{ (top) Responses at 2.23, 3.7 and
10.6~$\mu$m to a step increase of the central source from 1.4$\times
10^{12}$ to 1.8$\times 10^{12}L_{\sun}$ are shown for silicate grains
in the model described in the text. The grain density is assumed to
decrease with radius from the central source as (radius)$^{-2}$.
(bottom) Expected 10--90\% rise times are shown for the observations at
2.23, 3.7 and 10.6~$\mu$m if, in the model described in the text, the
luminosity of the central source increases by 50\%. The central
luminosities included in the figure span the  luminosity range of the
quasars in the sample  and the predictions  are shown for both silicate
(solid lines) and graphite grains (dashed lines) in the simple thermal
model described in the text.}
\figcaption[neugebauer.fig10.ps]{ (top) A histogram is given of the 
median variability, as
defined in the text, of the 10.6~$\mu$m observations in this
study. The sample is divided into radio quiet (single hatches) and
radio loud (cross hatches) quasars.
\newline
(bottom) A comparison is shown of the median variability measured
at 10.6~$\mu$m in this study (cross hatched bars) 
and that measured in the B band (0.44~$\mu$m) by Netzer et al. (1996) for
a sample of radio loud quasars (single hatched bars).} 
  
\figcaption[neugebauer.fig11.ps]{ The average structure functions for
the 10.6~$\mu$m observations of all 25 of the quasars in this sample
are shown.  The vertical error bars assigned to each point on a
structure function are the standard deviations in the means of each
bin, and thus reflect the spread in the estimates making up one bin,
while the error bars in the lags are arbitrarily sized to extend from
one to three quarters of the time interval between adjacent lag times.
The radio quiet quasars are distinguished from radio loud quasars as
indicated; the measurements of PG1226+023 were so much more numerous
than those of the other objects that they were treated separately.}
\figcaption[neugebauer.fig12.ps]{ The structure functions at all five 
wavelengths for
the quasar PG1226+023 are given.  The error bars have the same
meaning as those in Figure~11. The amplitude scale for each wavelength 
is different.}

\figurenum{13}
\figcaption[neugebauer.fig13.ps]{ The same as Figure 2, but for PG0007+106.}
\figurenum{14}
\figcaption[neugebauer.fig14.ps]{ The same as Figure 2, but for PG0026+129.}
\figurenum{15}
\figcaption[neugebauer.fig15.ps]{ The same as Figure 2, but for PG0050+124.}
\figurenum{16}
\figcaption[neugebauer.fig16.ps]{ The same as Figure 2, but for PG0157+001.}
\figurenum{17}
\figcaption[neugebauer.fig17.ps]{ The same as Figure 2, but for PG0804+761.}
\figurenum{18}
\figcaption[neugebauer.fig18.ps]{ The same as Figure 2, but for PG0844+349.}
\figurenum{19}
\figcaption[neugebauer.fig19.ps]{ The same as Figure 2, but for PG1119+120.}
\figurenum{20}
\figcaption[neugebauer.fig20.ps]{ The same as Figure 2, but for PG1211+143.}
\figurenum{21}
\figcaption[neugebauer.fig21.ps]{ The same as Figure 2, but for PG1229+204.}
\figurenum{22}
\figcaption[neugebauer.fig22.ps]{ The same as Figure 2, but for PG1351+640.}
\figurenum{23}
\figcaption[neugebauer.fig23.ps]{ The same as Figure 2, but for PG1402+261.}
\figurenum{24}
\figcaption[neugebauer.fig24.ps]{ The same as Figure 2, but for PG1411+442.}
\figurenum{25}
\figcaption[neugebauer.fig25.ps]{ The same as Figure 2, but for PG1426+015.}
\figurenum{26}
\figcaption[neugebauer.fig26.ps]{ The same as Figure 2, but for PG1440+356.}
\figurenum{27}
\figcaption[neugebauer.fig27.ps]{ The same as Figure 2, but for PG1613+658.}
\figurenum{28}
\figcaption[neugebauer.fig28.ps]{ The same as Figure 2, but for PG1617+175.}
\figurenum{29}
\figcaption[neugebauer.fig29.ps]{ The same as Figure 2, but for PG1634+706.}
\figurenum{30}
\figcaption[neugebauer.fig30.ps]{ The same as Figure 2, but for PG1700+518.}
\figurenum{31}
\figcaption[neugebauer.fig31.ps]{ The same as Figure 2, but for PG1704+608.}
\figurenum{32}
\figcaption[neugebauer.fig32.ps]{ The same as Figure 2, but for PG2130+099.}
\figurenum{33}
\figcaption[neugebauer.fig33.ps]{ The same as Figure 2, but for PG2302+029. 
In this figure only, the range of the ordinate scale has been truncated 
below the extremes of the 10.6~$\mu$m uncertainties.}


\normalsize
 
\begin{thebibliography}{4}
\bibitem[]{} Bahcall, J. N., Kirhakos, S., Saxe, D. H. \& Schneider, D.
P.  1997, \apj, 479, 642
\bibitem[]{} Bloom, S. D., Marscher, A. P., Gear, W. K., Teraesranta,
H., Valtaoja, E., Aller, H. D. \& Aller, M. F. 1994, \aj, 106, 398
\bibitem[]{} Bregman, J. N. 1994, in Multi-Wavelength Continuum
Emission of AGN, IAU Symposium 159, Vol. ed. Courvoisier, T. J.-L. and
Blecha, A.  (Geneva, Switzerland, Dordrecht, Boston, Kluwer, Academic
Press), p. 5
\bibitem[]{} Courvoisier, T. J., Robson, E. I., Blecha, A., Bouchet,
P., Hughes, D. H., Krisciunas, K. \& Schwarz, H. E. 1988, \nat, 335,
330
\bibitem[]{} Cutri, R. M., Wisniewski, W. Z., Rieke, G. H. \& Lebofsky,
M. J.  1985, \apj, 296, 423
\bibitem[]{} Draine, B. T. \& Lee, H. M. 1984, \apj, 285, 89
\bibitem[]{} Elias, J. H., Frogel, J. A., Matthews, K. \& Neugebauer,
G. 1982, \aj, 87, 1029
\bibitem[]{} Elvis, M.S. et al.1994, \apj, 95, 1
\bibitem[]{}Hoffleit, D. 1964, Catalogue of Bright Stars (New Haven, 
Connecticut, Yale University Observatory)
\bibitem[]{} Hughes, P. A., Aller, H. D. \& Aller, M. F. 1992, \apj,
396, 469
\bibitem[]{}Kukula, M. J., Dunlop, J. S., Hughes, D. H. \& Rawlings, S. 
1998, \mnras, 297, 366
\bibitem[]{} Kellermann, K. I., Sramek, R. A., Schmidt, M., Green, R.
F. \& Shaffer, D. B. 1994, \aj, 108, 1163
\bibitem[]{} Mathis, J. S., Rumpl, W. \& Nordsieck, K. H. 1977, \apj,
217, 425
\bibitem[]{} McLeod, K. K. \& Rieke, G. H. 1994a, \apj, 420, 58
\bibitem[]{} McLeod, K. K. \& Rieke, G. H. 1994b, \apj, 431, 137
\bibitem[]{} Netzer, H., et al. 1996, MNRAS, 279, 429
\bibitem[]{} Neugebauer, G., Green, R. F., Matthews, K., Schmidt, M.,
Soifer, B. T. \& Bennett, J. 1987, \apj, 63, S615
\bibitem[]{} Neugebauer, G., Matthews, K. \& Armus, L. 1995, \apj, 455,
L123
\bibitem[]{} Neugebauer, G., Soifer, B. T., Matthews, K. \& Elias, J.
H. 1989, \aj, 97, 957
\bibitem[]{} Press, W. H., Flannery, B. P., Teukolsky, S. A. \&
Vetterling, W.  T. 1985, Numerical Recipes The Art of Scientific
Computing Cambridge University Press)
\bibitem[]{} Press, W. H., Rybicki, G. B. \& Hewitt, J. N. 1992, \apj,
385, 404
\bibitem[]{} Rieke, G. H. \& Lebofsky, M. J. 1979, \araa, 17, 477
\bibitem[]{} Robson, E. I., et al. 1983, \nat, 305, 194
\bibitem[]{} Robson, E. I., et al. 1993, \mnras, 262, 249
\bibitem[]{} Sanders, D. B., Phinney, E. S., Neugebauer, G., Soifer, B.
T. \& Matthews, K. 1989, \apj, 347, 29
\bibitem[]{} Schmidt, M. \& Green, R. F. 1983, \apj, 269, 352
\bibitem[]{} Simonetti, J. H., Cordes, J. M. \& Heeschen, D. S. 1985,
\apj, 296, 46
\bibitem[]{} Smith, A. G., Nair, A. D., Leacock, R. J. \& Clements, S.
D. 1993, \aj, 105, 437
\bibitem[]{} Ulrich, M.-H., Maraschi, L. \& Urry, C. M. 1997, \araa,
35, 445
\bibitem[]{} Wilkes, B. J., Hooper, E. J., McLeod, K. K., Elvis, M. S.,
Impey, C. D., Lonsdale, C. J., Malkan, M. A. \& McDowell, J. C. 1998,
in The Universe as Seen by ISO, Vol. SP-427, Paris, ESA Special
Publication Series, preprint

\end{thebibliography}
\end{document}